\documentclass[12pt,a4paper]{article}
 \usepackage{epsf}
\begin{document}
\title{Signatures of the neutral top-pion in $e\gamma$ collisions }

\author{Chongxing Yue$^{(a,b)}$, Hong Li$^{b}$, Xuelei Wang $^{(a,b)}$   \\
 {\small a:CCAST (World Laboratory) P.O. BOX 8730. B.J. 100080 P.R. China}\\
 {\small b:College of Physics and Information Engineering,}\\
\small{Henan Normal University, Xinxiang  453002. P.R.China}
\thanks{This work is supported by the National Natural Science
Foundation of China(I9905004), the Excellent Youth Foundation of
Henan Scientific Committee(9911), and Foundation of Henan
Educational Committee.}
\thanks{E-mail:cxyue@public.xxptt.ha.cn} }
\date{}
\maketitle
\begin{abstract}
\hspace{5mm}We calculate the contributions of the neutral top-pion
$\pi_{t}^0$ to the process $e^-\gamma\rightarrow e^-\overline{t}c$
in the framework of topcolor-assisted technicolor(TC2) models and
discuss the possible of detecting $\pi_t^0$ at the high energy
linear $e^+e^-$ collider(LC). Our results show that $\pi_t^0$ can
give significant contributions to this process. With reasonable
values of the parameters in
 TC2 models, the cross section $\sigma$ can reach 0.19 fb which may be
detected at the $e\gamma$ collisions based on the future LC
experiments.
\end {abstract}


\newpage
 The mechanism of electroweak symmetry breaking (EWSB) remains
the most prominent mystery in elementary particle physics. Probing
 EWSB will be one of the most important tasks in the future high
energy colliders. Dynamical EWSB, such as technicolor (TC)
theory\cite{z1}, is an attractive idea that it avoids the
shortcomings of triviality and unnaturalness arising from the
elementary Higgs field. The simplest QCD-like TC model leads a too
large obligue correction to the electroweak parameters $ S $ and $
U $\cite{z2} and is already ruled out by the CERN $e^{+}e^{-}$
collider LEP precision electroweak measurement data \cite{z3}. To
solve the phenomenological difficulties of TC theories, TC2 models
\cite{z4} were proposed by combining TC interactions with topcolor
interactions for the third generation at the energy scale of about
1 TeV. TC2 theory is an attractive scheme in which there is an
explicit dynamical mechanism for breaking electroweak symmetry and
generating the fermion masses including the heavy top quark mass.
It is one of the important promising candidates for the mechanism
of EWSB.

In TC2 theory \cite{z4},  EWSB is driven mainly by TC
interactions, the extended technicolor (ETC) interactions give
contributions to all ordinary quark and lepton masses including a
very small portion of the top quark mass, namely
$m_{t}^{\prime}=\varepsilon m_t$ with a model-dependent parameter
$\varepsilon(\varepsilon\ll 1)$. The topcolor interactions also
make small contributions to EWSB and give rise to the main part of
the top quark mass $m_t-m_{t}^{\prime}=(1-\varepsilon)m_t$ similar
to the constituent masses of the light quarks in QCD. This means
that the associated top-pions $\pi_t^0,\pi_t^{\pm}$ are not the
longitudinal bosons W and Z, but separately, physically observable
objects. Thus top-pions can be seen as the characteristic feature
of TC2 theory. Studying the possible signatures of top-pions at
future high energy colliders can be used to test TC2 theory and
further probe the EWSB mechanism.

The virtual effects of the top-pions on the processes such as
$q\bar{q}\rightarrow t\bar{t}$, $q\bar{q^{\prime}}\rightarrow tb$,
$gg\rightarrow t\bar{t}(\bar{t}c)$, $e^+e^-\rightarrow tc\gamma
(Z)$, $\gamma \gamma \rightarrow t\bar{t}(\bar{t}c)$,
$t\rightarrow cv(v=\gamma, g, or Z)$, and $e\gamma \rightarrow
t\bar{b}\bar{\nu_e}$ have been studied in the literature, where
the signatures and observability of these new particles were
investigated in hadron colliders \cite{z5,z6}, $e^+e^-$ colliders
\cite{z7},$\gamma\gamma$ colliders \cite{z8} and $e\gamma$
colliders \cite{z9}. Ref.[9] has discussed the prospects of the
observation of the charged top-pions $\pi_t^{\pm}$ via the process
$e\gamma\rightarrow t\bar{b}\nu_e$ in $e\gamma$ colliders. In this
note, we calculate the contributions of the neutral top-pion
$\pi_t^0$ to the flavor changing neutral curent(FCNC)
 process $e^-\gamma\rightarrow e^{-}\bar{t}c$
and see whether $\pi_t^0$ can be detected via this process at
high-energy linear $e^+e^-$ colliders (LC) experiments. We find
that this process is important in probing the neutral top-pion.
With reasonable values of the parameters in TC2 models, the signal
rates can be fairly large, which may be detected at the $e\gamma$
colliders based on the LC experiments.

 For TC2 models \cite{z4},
the underlying interactions, topcolor interactions, are
non-universal and therefore do not posses a GIM mechanism. This is
an essential feature of this kind of models due to the need to
single out the top quark for condensation. This non-universal
gauge interactions result in the FC coupling vertices when one
writes the interactions in the quark mass eigenbasis. Thus the
top-pions predicted by this kind of models have large Yukawa
couplings to the third generation and can induce the FC scalar
couplings. The couplings of the neutral top-pion $\pi_t^0$ to the
ordinary fermions can be written as \cite{z4,z5}:
\begin{equation}
 \frac{m_t}{\sqrt{2}F_t}\frac{\sqrt{\nu_W^2-F_t^2}}{\nu_W}
 [k_{UR}^{tt}k_{UL}^{tt*}\bar{t_L}t_R\pi_t^0+\frac{m_b-m_{b}^{\prime}}{m_t}
 \bar{b_L}b_R\pi_t^0+k_{UR}^{tc}k_{UL}^{tt*}\bar{t_L}c_R\pi_t^0+h.c.],
\end{equation}
where $F_t\approx 50$ GeV is the top-pion decay constant,
$\nu_W=\nu/\sqrt{2}\approx 174$ GeV, and $m_b^{\prime}$ is the ETC
generated part of the bottom-quark mass. Similarly to Ref.[6], we
take $m_b^{\prime}=0.1\times \varepsilon m_t$. $k_{UL}^{tt}$ is
the matrix element of the unitary matrix $k_{UL}$ which the CKM
matrix can be derived as $V=k_{UL}^{-1}k_{DL}$ and $k_{UR}^{ij}$
are the matrix elements of the right-handed rotation matrix
$k_{UR}$. Their values can be written as:
\begin{equation}
k_{UL}^{tt}=1,\hspace{15mm}
k_{UR}^{tt}=1-\varepsilon,\hspace{15mm}
k_{UR}^{tc}\leq\sqrt{2\varepsilon-\varepsilon^2}.
\end{equation}
In the following calculation, we will take
$k_{UR}^{tc}=\sqrt{2\varepsilon-\varepsilon^2}$ and take the
parameter $\varepsilon$ as a free parameter.

The neutral top-pion $\pi_t^0$, as an isospin-triplet, can couple
to a pair of gauge bosons through the top quark triangle loop in
an isospin violating way similar to the couplings of QCD pion
$\pi^0$ to a pair of gauge bosons. For the top quark triangle
loop, the simple ABJ anomaly approach is not sufficient since the
top quark mass is only 175GeV. Here, we explicitly calculate the
top loop and obtain the following $\pi_t^0-\gamma-\gamma$
coupling:
\begin{equation}
-\frac{N_{C}\alpha_e(1-\varepsilon)m_t^2}{3\sqrt{2}\pi
F_t}C_0\pi_t^0
\epsilon_{\mu\nu\lambda\rho}(\partial^{\mu}A^{\nu})(\partial^{\lambda}
A^{\rho}),
\end{equation}
 where $N_C$ is the color index with $N_C=3$, $C_0=C_0(p_4, -p_4-p_3,
  m_t, m_t,m_t)$ is standard three-point scalar integral with $p_3$ and
  $p_4$ donating the momenta of the two incoming photons.

Ref.[4] has estimated the mass of the top-pion in the fermion loop
approximation and given 180 GeV $\leq m_{\pi_t}\leq $250 GeV for
$m_t=180$ GeV and $0.03\leq \varepsilon\leq 0.1$. Since the
negative top-pion corrections to the $Z\rightarrow b\bar{b}$
branching ratio $R_b$ become smaller when the top-pion is heavier,
the LEP/SLD data of $R_b$ give rise to certain lower bound on the
top-pion mass \cite {z10}. Ref.[11] has shown that the top-pion
mass is allowed to be in the range of a few hundred GeV depending
on the values of the parameters in TC2 models. Thus, at numerical
estimation, we take the mass of the $\pi_t^0$ to vary in range of
200 GeV-400 GeV in this letter. In this case, the possible decay
modes of $\pi_t^0$ are $\bar{t}c$, $b\bar{b}$, $gg$,
$\gamma\gamma$, $Z\gamma$ and $tt$ (if kinematically allowed).
Then we have
\begin{eqnarray}
\Gamma&=&\Gamma(\pi_t^0\rightarrow b\bar{b})+
\Gamma(\pi_t^0\rightarrow \bar{t}c)+\Gamma(\pi_t^0\rightarrow gg)+
\Gamma(\pi_t^0\rightarrow \gamma\gamma)+\Gamma(\pi_t^0\rightarrow
t\gamma) \nonumber \\ & & +\Gamma(\pi_t^0\rightarrow t\bar{t})
 (\hspace{2mm} for \hspace{2mm} m_{\pi_t}\geq 350
\hspace{2mm} GeV \hspace{2mm}).
\end{eqnarray}
In above equation, we have ignored the coupling of $\pi_t^0$ to a
pair of gauge bosons Z and taken $S_{\pi_t^{0}ZZ}\approx 0$.

The Feynman diagram for the neutral top-pion $\pi_t^0$
contributions to the process $e^-\gamma\rightarrow e^-\bar{t}c$ is
shown in Fig 1. With Eqs. (1)-(4), we can do the explicit
calculations of $\pi_t^0$ to the amplitude of the process
$e^-\gamma\rightarrow e^-\bar{t}c$.
\begin{eqnarray}
M&=&\frac{N_{c}\alpha_e(1-\varepsilon)^2m_t^2}{6\pi F_t^2}
\frac{\nu_W^2-F_t^2}{\nu_W^2}k_{UR}^{tc}C_0\bar{u}(p_c)\gamma_5v(p_t)
\frac{i}{p_{\pi}^2-m_{\pi}^2+im_{\pi}\Gamma}
\frac{-ig^{\alpha\beta}}{(p_{e^-}^{\prime}-p_{e^-})^2} \nonumber
\\ & &
\varepsilon_{\mu\nu\lambda\rho}(\partial^{\mu}\varepsilon_1^{\nu})
(\partial^{\lambda}\varepsilon_2^{\rho})\varepsilon_{\alpha}(p_{\gamma},
\lambda)\bar{u}(p_{e^-})\gamma_{\beta}u(p_{e^-})
\end{eqnarray}

The hard photon beam of the $e^-\gamma$ colliders can be obtained
from laser backscattering  at the LC \cite{z12}. We define that
$\sqrt{\hat{s}}$ and $\sqrt{s}$ are the center-of-mass energies of
the $e^-\gamma$ and $e^+e^-$ colliders, respectively. After
calculating the cross section $\sigma(\hat{s})$ for the subprocess
$e^-\gamma\rightarrow e^-\bar{t}c$, the total cross section
$\sqrt{s}$ at the LC experiments can be obtained by folding
$\sigma(\hat{s})$ with the backscattered laser photon spectrum
$f_{\gamma}(x) (\hat{s}=x^2s)$
 \begin{equation}
 \sigma = \int_{(m_t+m_c)/\sqrt{s}}^{x_{max}}dx\hat{\sigma}
 (\hat{s})f_{\gamma}(x).
\end{equation}
The backscattered laser photon spectrum $f_{\gamma}(x)$ is given
in Ref.[12]. Beyond a certain laser energy $e^+e^-$ pairs are
produced, which significantly degrades the photon beam. This leads
to a maximum $e\gamma$ centre of mass energy of $\sim 0.91\times
\sqrt{s}$.

In our calculation, we restrict the angles of the observed
particles relative to the beam, $\theta_{e^-}$ and $\theta_c$ to
the range $10^\circ\leq\theta_{e^-}$, $\theta_e\leq170^\circ$. We
further restrict the particle energy $E_e\geq 10$ GeV. For
simplicity, we have ignored the possible polarization for the
electron and photon beams. It has been shown \cite{z5} that the
neutral top-pion $\pi_t^0$ mainly couples to the right-handed
top$(t_R)$ or charm $(c_R)$ and the left-handed rotation element
$k_{UL}^{tc}$ is negligibly small. Thus, we only consider chiral
couplings of the $\pi_t^0$ to top-charm. Our results are all based
on right-handed couplings. To obtain numerical results, we take
$m_t=175$ GeV, $m_c=1.2$ GeV and $\alpha_e=\frac{1}{128}$
\cite{z13}. For estimating the number of the $e^-\bar{t}c$
 event, similarly to Ref.[9,14], we consider the $e^+e^-$
centre-of-mass energy $\sqrt{s}$ in the range of 300GeV-1500GeV
 appropriate to the TESLA/NLC/JLC high energy colliders and assume an
 integrated luminosity of $L=500 fb^{-1}$.

In Fig2, we show the cross section $\sigma$ of the process
$e^-\gamma\rightarrow e^-\bar{t}c$ as a function of the mass of
the neutral top-pion $m_{\pi_t}$ for the center-of-mass energy
$\sqrt{s}=500$ GeV and three values of the parameter
$\varepsilon$. One can see that neutral  top-pion $\pi_t^0$ can
give significant contributions to the process
$e^-\gamma\rightarrow e^-\bar{t}c$. We see from Fig.2 that the
cross section $\sigma$ is sensitive to the parameter
$\varepsilon$. The $\pi_t^0$ resonance contribution increases as
the parameter $\varepsilon$ increasing. This is because the total
decay width of $\pi_t^0$ decreases as $\varepsilon$
 increasing. The maximum value can reach 0.12fb for $\varepsilon=0.08$
and $m_{\pi_t}=270$GeV. Thus, there will be several tens of
$e^-\bar{t}c$ events to be generated which may be detected in the
future LC experiments.

To see the effect of the center-of-mass $\sqrt{s}$ on the
$\sigma$, we plot the $\sqrt{s}$ for $m_{\pi_t}$=250GeV and three
values of the parameter
 $\varepsilon$ in Fig.3. We can see from Fig.3 that the cross section
$\sigma$ is larger than $4\times10^{-2}fb$ for
$\sqrt{s}\geq500$GeV. For
 $m_{\pi_t}=250$GeV and $0.03\leq\varepsilon\leq0.08$, the maximum value can
reach 0.19fb. In this case, there are about 90 $\bar{t}c$ events
to be generated in the future LC experiments.

TC2 models also predict the existence of the neutral CP-even sate,
called top-Higgs boson $h_t^0$. The main difference between
$\pi_t^0$ and $h_t^0$ is that $h_t^0$ can couple to gauge boson
pairs $WW$ and $ZZ$ at tree-level, which is similar to that of the
standard model(SM) Higgs boson $H^0$\cite{z5}. However, the
coupling coefficients of the couplings $h_t^0WW$ and $h_t^0ZZ$
 are suppressed by the factor $\frac{F_t}{\nu_W}$ with respect to that of
 $H^0$. Thus, the contributions of the top-Higgs boson $h_t^0$ to the
 process $e^-\gamma\rightarrow e^-\bar{t}c$ are similar to that of
$\pi_t^0$. In the most of the parameter space of TC2 models, the
cross section $\sigma$ contributed by the top-Higgs $h_t^0$ is
also in the range of $10^{-1}-10^{-2} fb$.

The search for FCNC processes is one of the most interesting
possibilities to test the SM, with the potential for either
 discovering or putting stringent bounds on new physics. In the SM, there
are no FCNC at tree-level and at one-loop level they are GIM
suppressed. In models beyond SM, new particles may appear in the
loop and have significant contributions to the FCNC processes.
Therefore, the processes can give an ideal place to search the
signals of the new particles. In this letter, we calculated the
contributions of the neutral top-pion $\pi_t^0$ to the FCNC
process $e^-\gamma\rightarrow e^-\bar{t}c$ in the framework of TC2
models and disscused the possible of detecting this new particle
in the future LC experiments. Our numerical results show that the
cross section $\sigma$
 given by $\pi_t^0$ is in the range of the $10^{-1}-10^{-2}fb$. With
 reasonable values of the parameters, the cross section $\sigma$ can reach
 to $0.19fb$.
 So it is possible to detect the signals of the neutral top-pion $\pi_t^0$
 via the process $e^-\gamma\rightarrow e^-\bar{t}c$ at the
 $e\gamma$ colliders based on the LC experiments.

\newpage
\vskip 2.0cm
\begin{center}
{\bf Figure captions}
\end{center}
\begin{description}
\item[Fig.1:]Feynman diagram for contributing to the process
$e^-\gamma\rightarrow e^-\bar{t}c$ from the neutral top-pion
$\pi_t^0$.
\item[Fig.2:]The cross section $\sigma$ as a function of $m_{\pi_t}$
for the center-of-mass energy $\sqrt{s}=500$ GeV and
$\varepsilon=0.03$ (solid line), 0.05 (dotted line) and 0.08
(dashed line).
\item[Fig.3:]The cross section $\sigma$ as a function of
$\sqrt{s}$ for the parameter $m_{\pi_t}=250$GeV and
$\varepsilon=0.03$ (solid line), 0.05 (dotted line) and 0.08
(dashed line).

\end{description}

\newpage
\begin{figure}[pt]
\begin{center}
\begin{picture}(250,200)(0,0)
\put(20,-180){\epsfxsize120mm\epsfbox{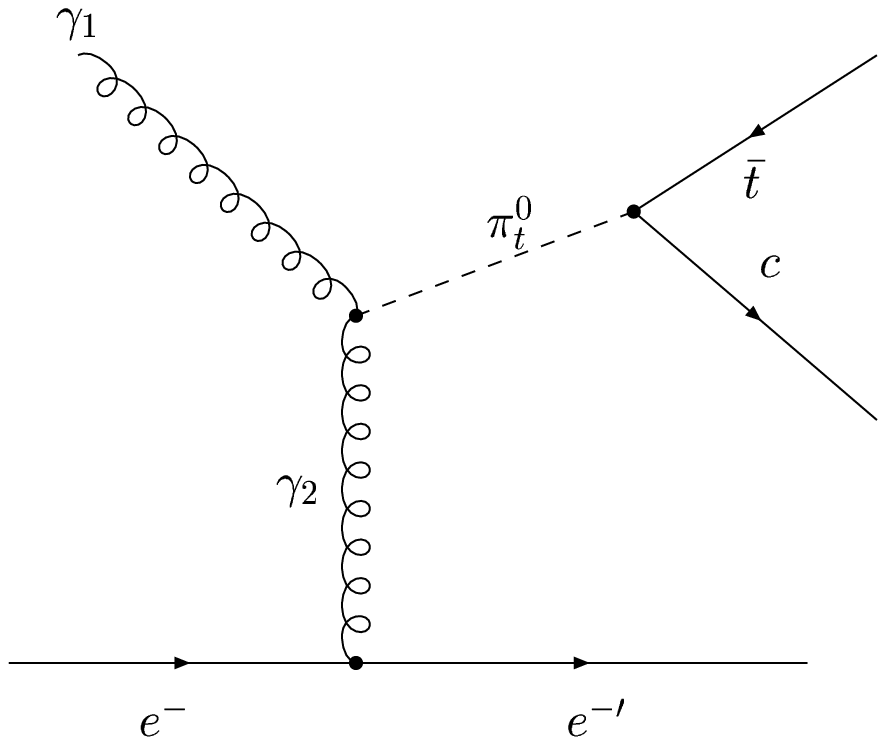}}
 \put(120,-10){Fig.1}
\end{picture}
\end{center}
\end{figure}

\begin{figure}[hb]
\begin{center}
\begin{picture}(250,200)(0,0)
\put(-50,-80){\epsfxsize120mm\epsfbox{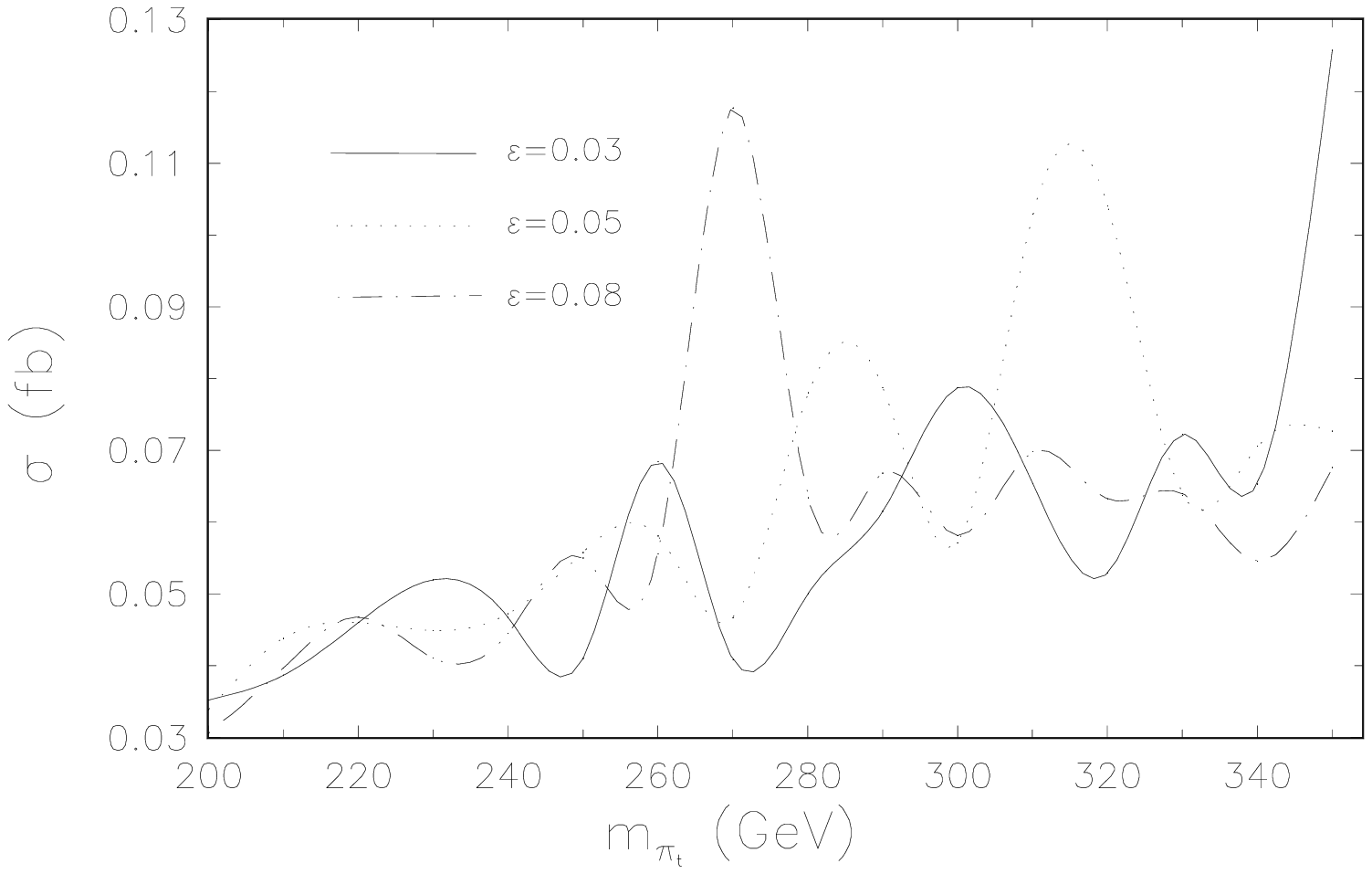}}
 \put(130,-120){Fig.2}
\end{picture}
\end{center}
\end{figure}

\begin{figure}[hb]
\begin{center}
\begin{picture}(250,200)(0,0)
\put(-50,10){\epsfxsize120mm\epsfbox{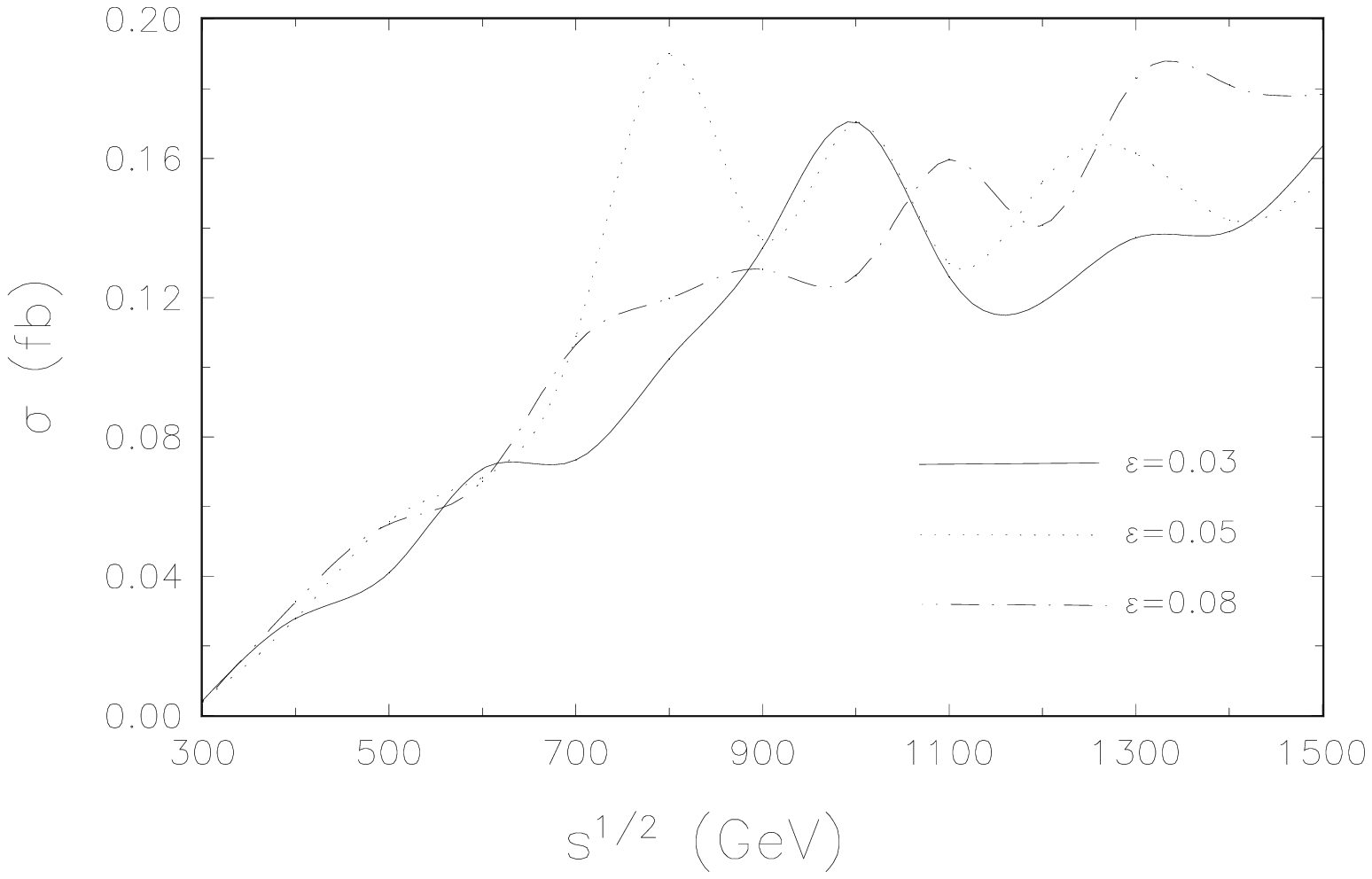}}
 \put(120,-40){Fig.3}
\end{picture}
\end{center}
\end{figure}


\newpage
\begin{thebibliography}{99}
\bibitem{z1}S. Weinberg, {\em Phys. Rev. D}{\bf 13}(1976)974;
{\em D}{\bf 19}(1997)1277; L. Susskind, {\em ibid. D}{\bf
20}(1979)2619; S. Dimopoulos and L. Susskind, {\em Nucl. Phys.
B}{\bf 155}(1979)237; E. Eichten and K. Lane, {\em Phys. Lett.
B}{\bf 90}(1980)125.
\bibitem{z2}M. Peskin and T. Takeuchi, {\em Phys. Rev. Lett. }{\bf
65}(1990)964.
\bibitem{z3}J. Erler and P. Lanagaclar, {\em  Review of Particle
Physics, Eur. Phys. J. C}{\bf 3} (1998)90;
 K. Hagiwara, D. Haidt, and S. Matsumoto, {\em Eur. Phys.
J. C}{\bf 2}(1998)95.
\bibitem{z4}C. T. Hill, {\em Phys. Lett. B}{\bf 345}(1995)483;
K. Lane and F. Eichten, {\em Phy. Lett. B.}{\bf 352}(1995)382; K.
Lane, {\em Phys. Lett. B}{\bf 433}(1998)96; G. Cvetic, {\em Rev.
Mod. Phys. }{\bf 71}(1999)513.
\bibitem{z5}Hong-Jian He and C. P. Yuan, {\em Phys. Rev. Lett.
}{\bf 83}(1999)83; G. Burdman, {\em Phys. Rev. Lett. }{\bf
83}(1999)2888.
\bibitem{z6}Chongxing Yue, et al., {\em Phys. Rev. D}{\bf 55}(1997)5541;
Gongru Lu, Chongxing Yue
and Jinshu Huang, {\em Phys. Rev. D}{\bf 57}(1998)1755.
\bibitem{z7}Chongxing Yue, et al., {\em Phys. Rev. D}{\bf 63}(2001)115002.
\bibitem{z8}Hongyi Zhou, et al., {\em Phys. Rev. D }{\bf 57}(1998)4205;
Chongxing Yue, et al., {\em Phys. Lett. B} {\bf 496}(2000)93.
\bibitem{z9}Xuelei Wang, et al., {\em Phys. Rev. D}{\bf 60}(1999)014002.
\bibitem{z10}G. Burdman and D. Kominis, {\em  Phys. Lett. B}
{\bf 403}(1997)101; W. Loinaz and T. Takuchi, {\em Phys. Rev.
D}{\bf 60}(1999)015005.
\bibitem{z11}Chongxing Yue, et al., {\em Phys. Rev. D }{\bf 62}(2000)055005.
\bibitem{z12}G. Jikia, Nucl. {\em Phys. B}{\bf 374}(1992)83; O.
J. P. Eholi, et al., {\em Phys. Rev. D}{\bf 47}(1993)1889; K. M.
Cheuny, {\em Phys. Rev. D}{\bf 47}(1993)3750.
\bibitem{z13}Paratical Data Group, {\em Eur. Phys. J. C}{\bf
15}(2000)1.
\bibitem{z14}S. Godfrey, P. Kalyniak and N. RomanenKo, hep-ph/0108285;
hep-ph/0100191.
\end{thebibliography}
\end{document}